\documentclass[aps,prb,twocolumn,amsmath,amssymb,showpacs]{revtex4}

\usepackage{graphicx}
\usepackage{graphics}
\usepackage{dcolumn}
\usepackage{bm}

\begin{document}

\title{Conductance characteristics between a normal metal and a
two-dimensional Fulde-Ferrell-Larkin-Ovchinnikov superconductor: the
Fulde-Ferrell state}

\author{Qinghong Cui$^1$, C.-R. Hu$^2$, J.Y.T. Wei$^3$, and Kun Yang$^1$}

\affiliation{$^1$National High Magnetic Field Laboratory and Department of
Physics, Florida State University, Tallahassee, Florida 32306, USA}
\affiliation{$^2$Department of Physics, Texas A\&M University, College Station,
Texas 77843, USA}
\affiliation{$^3$Department of Physics, University of Toronto, 60 St. George St.
Toronto, ON M5S1A7, Canada}

\date{\today}

\begin{abstract}

The Fulde-Ferrell-Larkin-Ovchinnikov (FFLO) state has received
renewed interest recently due to the experimental indication of its
presence in CeCoIn$_5$, a quasi 2-dimensional (2D) $d$-wave
superconductor. However direct evidence of the spatial variation of
the superconducting order parameter, which is the hallmark of the
FFLO state, does not yet exist. In this work we explore the possibility of
detecting the phase structure of the order parameter directly using
conductance spectroscopy through micro-constrictions, which probes the phase
sensitive surface Andreev bound states of $d$-wave superconductors. We
employ the Blonder-Tinkham-Klapwijk formalism to calculate the
conductance characteristics between a normal metal (N) and a 2D $s$-
or $d_{x^2-y^2}$-wave superconductor in the Fulde-Ferrell state, for all
barrier parameter $z$ from the point contact limit ($z=0$) to the tunneling
limit ($z \gg 1$). We find that the zero-bias conductance peak due to these
surface Andreev bound states observed in the uniform $d$-wave superconductor is
split and shifted in the Fulde-Ferrell state. We also clarify what
weighted bulk density of states is measured by the conductance in
the limit of large $z$.

\end{abstract}


\maketitle

\section{Introduction}

In the early 1960's, Fulde and Ferrell~\cite{FF1964} and Larkin and
Ovchinnikov~\cite{LO1965} proposed the possibility that a
superconducting state with a periodic spatial variation of the order
parameter would become stable when a singlet superconductor is
subject to a large Zeeman splitting. The Zeeman splitting could be
due to either a strong magnetic field
or an internal exchange field. Under such a strong magnetic or
exchange field, there is a splitting of the Fermi surfaces of
spin-up and -down electrons, and condensed pairs of electrons with
opposite spins across the Fermi surface may be formed to lower the
free energy from that of a normal spin-polarized state. These pairs
have a non-zero total momentum 2{\bf q}, which causes the phase of
the superconducting order parameter to vary spatially with the wave
number 2{\bf q}. This state is known as the Fulde-Ferrell (FF)
state. Larkin and Ovchinnikov (LO), on the other hand, proposed
independently an alternative scenario, in which the order parameter
is real, but varies periodically in space, possibly in more than one
directions. Both types of states are now known (collectively) as the
Fulde-Ferrell-Larkin-Ovchinnikov (FFLO) state. It has not yet been
observed in conventional low-$T_c$ $s$-wave superconductors, which
are mostly three-dimensional, probably because the orbital effect of
the magnetic field dominates the Zeeman effect. The situation has
been changed by experimental results suggestive of the FFLO state in
heavy-fermion, quasi-1D organic, or high-$T_c$
superconductors;~\cite{gloos, tachiki, modler1996a, modler1996b,
brooks, singleton, mazin, tanatar, annett} many of these compounds are quasi-one or
two-dimensional, thus the orbital effect is weak when the magnetic
field is in the conducting plane or along the chain. Recent
experimental results in CeCoIn$_5$, a quasi-2D $d$-wave
superconductor, are particularly encouraging.~\cite{radovan,
bianchi2003, capan, watanabe, martin, kakuyanagi, bianchi2005} We
note in passing that this subject is also of interest to the nuclear and
particle physics communities because of the possible realization of
the FFLO state in high density quark matter and nuclear
matter,~\cite{casalbuoni} as well as in cold fermionic atom
systems.\cite{coldatom}

The most unambiguous evidence of the FFLO states should be based on
phase-sensitive experiments that can directly reveal the spatial
variation of the phase or sign of the order parameter. One
possibility is the Josephson effect.~\cite{kun2000} To the best of
our knowledge there has been no report on this or other phase
sensitive measurements thus far. In this paper, we consider an
alternative phase sensitive probe, i.e., the conductance
spectroscopy through a micro-constriction.
A powerful method to calculate the differential
conductance [$G(V) \equiv dI(V)/dV$] characteristics of a normal
metal/superconductor junction (NSJ) was developed by Blonder, Tinkham, and
Klapwijk~\cite{BTK}, which unified quasiparticle tunneling and
Andreev reflection~\cite{andreev, deutscher}.
Only $s$-wave superconductors were considered in that work, but
the method has since been generalized to the $d$-wave
case.~\cite{hu1994, tanaka}
When the sign of the order parameter experienced by the electron- and hole-like
quasiparticles (QPs) before and after specular reflection at the
junction interface changes, zero-energy Andreev bound states (ABSs) (also
called the midgap states~\cite{hu1994}) are formed at the S side of
the N/S interface (for $z\ne 0$). These ABSs will give rise to a zero-bias
conductance peak (ZBCP) in the tunneling spectrum of the NSJ. A well known
example of such NSJ is when the junction interface along the nodal line
of the $d_{x^2-y^2}$-wave superconductor
[(110) contact].~\cite{hu1994, tanaka, yang, xu, kashiwaya, lofwander}
This feature has since been widely used to identify the order-parameter
symmetry of unconventional superconductors.~\cite{alff, fogelstrom, wei1998,
laube, liu, biswas, wei2005} We would like to emphasize here that the ABSs are
consequences of the phase change of the $d$-wave order
parameter~\cite{harlingen, tsuei}, and
thus their spectra should also be sensitive to the spatial variation
of the order parameter. In this paper, we will show this sensitivity by explicit
calculations, and from their spectra detected via conductance spectroscopy through
a micro-constriction, one can extract the momentum of the
superconducting order parameter. In the present work, we will focus
on the FF state; the more general LO cases are currently being
investigated and will be presented elsewhere.~\cite{note1} As we will
show below, for a $d_{x^2-y^2}$-wave superconductor in the FF state, the ZBCP
observed in a (110) contact is split and
shifted by both the Zeeman field and pair momentum; the
latter can be determined from the splitting.

This paper is organized as follows. In section II we introduce the
model and present the self-consistent mean-field solutions for both
$s$- and $d_{x^2-y^2}$-wave superconductors in the FF phases. The electron density
of states (DOS) of the FF states is also obtained. The scheme to
calculate the conductance characteristics is presented in section
III, with numerical results for both $s$-wave and $d_{x^2-y^2}$-wave cases, and their
relation with the corresponding bulk electron DOS is discussed. The
features associated with the ABSs in the FF phase are explored in
section IV, where we solve their spectra in the FF state. We present
further discussion on our results and a summary in section V. In this work, we consider zero temperature only, and calculations on $d$-wave superconductors are simply referred to $d_{x^2-y^2}$-wave.


\section{The Fulde-Ferrell state and bulk density of states}

We begin with the mean field Hamiltonian for a 2D superconductor in the BCS state
($\mathbf{q}=0$) or the FF state ($\mathbf{q}\ne 0$):
\begin{eqnarray} \label{eq:hmltmf}
\lefteqn{\mathcal{H}_{\mathrm{MF}} = \sum_{\mathbf{k} \sigma}
   (\xi_{\mathbf{k}} + \sigma \mu_0 B) c_{\mathbf{k} \sigma}^{\dagger}
   c_{\mathbf{k} \sigma} {} } \nonumber \\
&& {}- \sum_{\mathbf{k}}(\Delta_{\mathbf{kq}} c_{\mathbf{k+q}\uparrow}^{\dagger}
   c_{\mathbf{-k+q}\downarrow}^{\dagger} + \Delta_{\mathbf{kq}}^{\ast}
   c_{\mathbf{-k+q}\downarrow} c_{\mathbf{k+q}\uparrow}),
\end{eqnarray}
where $\xi_{\mathbf{k}}$ is the single-particle kinetic energy
relative to the Fermi energy $\epsilon_F$, $B$ is the Zeeman
magnetic field, $\mu_0$ is the magnetic moment of the electron, $\Delta_{\mathbf{kq}}$
is the pairing potential and satisfies the self-consistent condition,
\begin{equation} \label{eq:self}
\Delta_{\mathbf{kq}} = - \sum_{\mathbf{k'}} V_{\mathbf{kk'}}
   \langle c_{\mathbf{-k'+q}\downarrow} c_{\mathbf{k'+q}\uparrow} \rangle .
\end{equation}
For $s$-wave, we have
$V_{\mathbf{kk'}}=-V_0$, while for $d$-wave, we have
$V_{\mathbf{kk'}}=-V_0 \cos(2 \theta_{\mathbf{k}}) \cos(2
\theta_{\mathbf{k'}})$ (here, $|\xi_{\mathbf{k}}|, \,
|\xi_{\mathbf{k'}}| < \hbar \omega_c \ll \epsilon_F$, $\omega_c$ is
the cutoff, $\theta_{\mathbf{k}}$ is the azimuthal angle of {\bf
k}). The orbital coupling between the magnetic field and electron motion is
absent when the field is in the plane.
The mean field Hamiltonian could be rewritten as
\begin{equation}
\mathcal{H}_{\mathrm{MF}} = \sum_{\mathbf{k}}
   (c_{\mathbf{k+q} \uparrow}^{\dagger}, \, c_{\mathbf{-k+q} \downarrow})
   \hat{\mathrm{H}}_{\mathbf{k}}\left(\begin{array}{c}
   c_{\mathbf{k+q}\uparrow} \\ c_{\mathbf{-k+q} \downarrow}^{\dagger}
   \end{array} \right) + \mathrm{const.}
\end{equation}
where
\begin{displaymath}
\hat{\mathrm{H}}_{\mathbf{k}} = \left( \begin{array}{cc}
   \xi_{\mathbf{k+q}} + \mu_0 B & -\Delta_{\mathbf{kq}} \\
   -\Delta_{\mathbf{kq}}^{\ast} & -\xi_{\mathbf{k-q}} + \mu_0 B
\end{array} \right).
\end{displaymath}
To diagonalize it, we perform the Bogoliubov-Valatin transformation
\begin{equation} \label{eq:qptrans1}
\left(\begin{array}{c}
   c_{\mathbf{k+q}\uparrow} \\ c_{\mathbf{-k+q}\downarrow}^{\dagger}
\end{array} \right)
= \left(\begin{array}{cc}
   u_{\mathbf{k}}^{\ast} & v_{\mathbf{k}} \\
   -v_{\mathbf{k}}^{\ast} & u_{\mathbf{k}}
\end{array}\right)
\left(\begin{array}{c}
   \alpha_{\mathbf{k}} \\ \beta_{\mathbf{k}}^{\dagger}
\end{array}\right),
\end{equation}
and choose
\begin{eqnarray*}
&& \frac{u_{\mathbf{k}}}{v_{\mathbf{k}}} =
   \frac{\xi_{\mathbf{k}}^{(s)} + E_{\mathbf{k}}}{\Delta_{\mathbf{kq}}}, \\
&& |u_{\mathbf{k}}|^2 = \frac{1}{2}
   \big(1 + \frac{\xi_{\mathbf{k}}^{(s)}}{E_{\mathbf{k}}}\big)
   = 1 - |v_{\mathbf{k}}|^2,
\end{eqnarray*}
where $E_{\mathbf{k}} = \sqrt{\Delta_{\mathbf{kq}}^2 +
\xi_{\mathbf{k}}^{(s)2}}$, $\xi_{\mathbf{k}}^{(s)} =
(\xi_{\mathbf{k+q}} + \xi_{\mathbf{k-q}})/2$, and
$\xi_{\mathbf{k}}^{(a)} = (\xi_{\mathbf{k+q}} -
\xi_{\mathbf{k-q}})/2$, from which we get the diagonalized
Hamiltonian
\begin{equation}
\mathcal{H}_{\mathrm{MF}} = \sum_{\mathbf{k}} (E_{\mathbf{k} +}
   \alpha_{\mathbf{k}}^{\dagger} \alpha_{\mathbf{k}} + E_{\mathbf{k} -}
   \beta_{\mathbf{k}}^{\dagger} \beta_{\mathbf{k}}) + \mathrm{const.}
\end{equation}
with eigenenergies ($\sigma = \pm 1$)
\begin{equation}
E_{\mathbf{k} \sigma} = E_{\mathbf{k}} +
   \sigma (\mu_0 B + \xi_{\mathbf{k}}^{(a)}).
\end{equation}
There are regions in the {\bf k}-space where the Cooper pairs are
destroyed and occupied by electrons of one spin species; these are
states with $E_{\mathbf{k}\sigma} < 0$. In these cases the
Bogoliubov-Valatin transformation (Eq.~(\ref{eq:qptrans1})) should
be replaced by
\begin{equation} \label{eq:qptrans2}
\left(\begin{array}{c}
   c_{\mathbf{k+q}\uparrow} \\ c_{\mathbf{-k+q}\downarrow}^{\dagger}
\end{array} \right)
= \left(\begin{array}{cc}
   u_{\mathbf{k}}^{\ast} & v_{\mathbf{k}} \\
   -v_{\mathbf{k}}^{\ast} & u_{\mathbf{k}}
\end{array}\right)
\left(\begin{array}{c}
   \alpha_{\mathbf{k}}^{\dagger} \\ \beta_{\mathbf{k}}^{\dagger}
\end{array}\right)
\end{equation}
when $E_{\mathbf{k}+} < 0$, or
\begin{equation} \label{eq:qptrans3}
\left(\begin{array}{c}
   c_{\mathbf{k+q}\uparrow} \\ c_{\mathbf{-k+q}\downarrow}^{\dagger}
\end{array} \right)
= \left(\begin{array}{cc}
   u_{\mathbf{k}}^{\ast} & v_{\mathbf{k}} \\
   -v_{\mathbf{k}}^{\ast} & u_{\mathbf{k}}
\end{array}\right)
\left(\begin{array}{c}
   \alpha_{\mathbf{k}} \\ \beta_{\mathbf{k}}
\end{array}\right)
\end{equation}
when $E_{\mathbf{k}-} < 0$.~\cite{note} Then, the diagonalized Hamiltonian is
\begin{equation}
\mathcal{H}_{\mathrm{MF}} = \sum_{\mathbf{k}} (|E_{\mathbf{k}+}|
   \alpha_{\mathbf{k}}^{\dagger} \alpha_{\mathbf{k}} + |E_{\mathbf{k}-}|
   \beta_{\mathbf{k}}^{\dagger} \beta_{\mathbf{k}}) + \mathrm{const.}
\end{equation}
with positive quasiparticle energies.

\begin{figure}
\includegraphics[width = 0.35\textwidth]{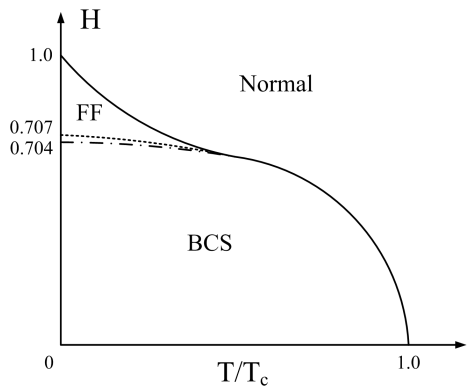}
\caption{\label{fig:phase1} Qualitative sketch of the phase
diagram of an $s$-wave superconductor under a Zeeman field, including the
possibility of Fulde-Ferrell state. Solid line shows the phase
transition between the superconducting state and the normal spin
polarized state; dash-dotted line indicates the phase transition
between the usual BCS state and the FF state; dotted line gives the
Clogston-Chandrasekhar critical field.}
\end{figure}

\begin{figure}
\includegraphics[width = 0.35\textwidth]{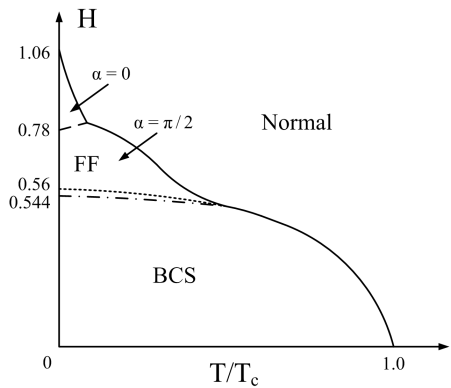}
\caption{\label{fig:phase2} Qualitative sketch of the phase
diagram of a $d$-wave superconductor under a Zeeman field, including the possibility
of Fulde-Ferrell state. The curves have the same meaning as in
Fig.~\ref{fig:phase1}, except that the dashed line is the phase
boundary separating FF states with {\bf q} along the antinodal
direction and the nodal direction. Here, $\alpha$ is the angle
between {\bf q} and the antinodal direction.}
\end{figure}

In the weak coupling limit, the total energy of the system is given by
\begin{equation} \label{eq:FFenergy}
\langle \mathcal{H}-\mu N \rangle = \sum_{\mathbf{k}} \left\{ \begin{array}{ll}
   2\xi_{\mathbf{k}}^{(s)} |v_{\mathbf{k}}|^2, & E_{\mathbf{k}, \pm 1}>0 \\
   \xi_{\mathbf{k+q}} + \mu_0 B, & E_{\mathbf{k} +} < 0 \\
   \xi_{\mathbf{k-q}} - \mu_0 B, & E_{\mathbf{k} -} < 0
\end{array} \right\} - \frac{\Delta_q^2}{V_0}.
\end{equation}
Here, for an $s$-wave superconductor, we have $\Delta_{\mathbf{kq}} = \Delta_q$,
and for a $d$-wave
superconductor, we have $\Delta_{\mathbf{kq}} = \Delta_q \cos 2\theta_{\mathbf{k}}$.
For a given Zeeman field $B$, we calculate
the energy of the pairing state numerically using Eq.~(\ref{eq:FFenergy}) with
$Q = q v_F/\Delta_0$ varied, and compare it with that of the normal spin polarized
state, in order to find the ground state. In our
numerical calculation presented below, we take $\hbar\omega_c = 10
\Delta_0$ (here $\Delta_0$ is the gap of the usual BCS state at
$T=0$ without a Zeeman field, and for $d$-wave it is the gap
along antinodal direction or maximum gap), and $H = \mu_0 B/\Delta_0$.
The calculation of the self-consistent
pairing potential shows that the superconducting state is destroyed at $H>1$
for $s$-wave~\cite{shimahara1994} and at $H>1.06$ for $d$-wave.~\cite{maki1996,kun1998}
In the superconducting regime, for $s$-wave case, at
$H_{p1} \approx 0.704$, a transition occurs and the FF state becomes the
ground state. For $d$-wave case, the critical Zeeman field is
$H_{p1} \approx 0.544$ where an FF phase occurs with {\bf q} along the nodal
direction. At a higher $H_{p1}' \approx 0.78$, the FF state with {\bf q}
along the antinodal direction dominates.~\cite{kun1998} The critical fields $H_{p1}$
are slightly smaller than the Clogston-Chandrasekhar fields
which is $H_{p1} = 1/\sqrt{2}$ for $s$-wave~\cite{clogston}, and
$\approx 0.56$ for $d$-wave~\cite{kun1998} (see also Ref.
\onlinecite{vorontsov}). The qualitative illustrations of the phase
diagrams are presented in Figs.~\ref{fig:phase1} and
\ref{fig:phase2}.

The finite momentum of the order parameter leads to a nonzero
supercurrent in the ground state. However, the supercurrent is
expected to be compensated by the backflow of the quasiparticle
current, and therefore, we have a zero total current.~\cite{FF1964} Here we
demonstrate that the total current is indeed zero for the self-consistent
mean-field ground state. The total current operator can be written
as
\begin{equation} \label{eq:current}
\mathbf{J} = \frac{e}{m} \sum_{\mathbf{k}} ((\mathbf{k+q})
   c_{\mathbf{k+q} \uparrow}^{\dagger} c_{\mathbf{k+q} \uparrow}
   -(\mathbf{k-q}) c_{\mathbf{-k+q} \downarrow}^{\dagger}
   c_{\mathbf{-k+q} \downarrow})
\end{equation}
and in the superconducting state, its expectation value is
\begin{equation} \label{eq:FFcurrent}
\langle \mathbf{J} \rangle = -\frac{e}{m} \sum_{\mathbf{k}}
\left\{ \begin{array}{ll}
   2\mathbf{q} |v_{\mathbf{k}}|^2, & E_{\mathbf{k}, \pm 1}>0 \\
   \mathbf{q+k}, & E_{\mathbf{k} +} < 0 \\
   \mathbf{q-k}, & E_{\mathbf{k} -} < 0\,.
\end{array} \right\}.
\end{equation}
By differentiate Eq.~(\ref{eq:FFenergy}) at fixed $\Delta$ we can
verify that
\begin{displaymath}
\langle \mathbf{J} \rangle = -e \frac{\partial}{\partial \mathbf{q}}
   \langle \mathcal{H}-\mu N \rangle\,.
\end{displaymath}
Since $\langle \mathcal{H}-\mu N \rangle$ is minimized in the ${\bf
q},\Delta$ space, the current must be zero.

\begin{figure}
\includegraphics[width = 0.35\textwidth]{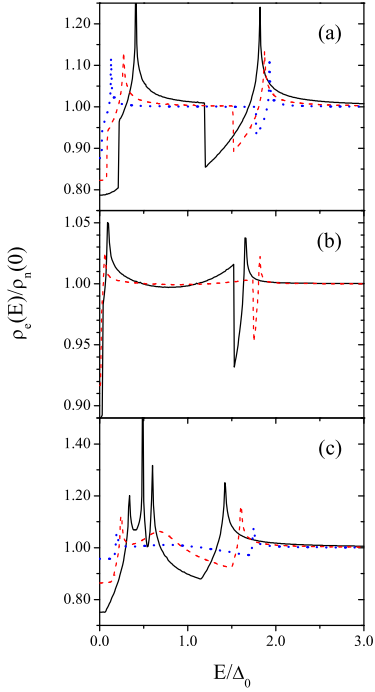}
\caption{\label{fig:dos1} Representative electron density of states
for various Zeeman fields and both $s$- and $d$-wave superconductors. (a) $s$-wave: solid, $H=0.704$, $Q=0.804$; dash, $H=0.8$, $Q=0.892$; dot, $H=0.9$,
$Q=0.952$. (b) $d$-wave
with {\bf q} along the antinodal direction: solid, $H = 0.78$, $Q=0.808$;
dash, $H = 0.88$, $Q=0.9$. (c) $d$-wave with {\bf q} along the nodal
direction: solid, $H = 0.544$, $Q=0.608$; dash, $H = 0.68$, $Q=0.816$;
dot, $H = 0.776$, $Q=0.94$.}
\end{figure}

The electron DOS can be evaluated
using
\begin{eqnarray} \label{eq:dos0}
\lefteqn{\rho(E) = \sum_{\mathbf{k}, E_{\mathbf{k} \pm} > 0} |u_{\mathbf{k}}|^2
   \big( \delta(E - E_{\mathbf{k}+}) + \delta(E - E_{\mathbf{k}-}) \big) {}}
   \nonumber \\
&& {} + \sum_{\mathbf{k}, E_{\mathbf{k} +} < 0} \big(|v_{\mathbf{k}}|^2
   \delta(E + E_{\mathbf{k}+}) + |u_{\mathbf{k}}|^2 \delta(E - E_{\mathbf{k}-})
   \big) \nonumber \\
&& {} + \sum_{\mathbf{k}, E_{\mathbf{k} -} < 0} \big(|u_{\mathbf{k}}|^2
   \delta(E - E_{\mathbf{k}+}) + |v_{\mathbf{k}}|^2 \delta(E + E_{\mathbf{k}-})
   \big). \nonumber \\
\end{eqnarray}
Within the approximation of $\xi_{\mathbf{k}}^{(s)} \approx
\xi_{\mathbf{k}}$ and $\xi_{\mathbf{k}}^{(a)} \approx q v_F
\cos(\theta_{\mathbf{k}} - \theta_{\mathbf{q}})$, it can be seen
that for an arbitrary state {\bf k}, there is always another state
$\mathbf{k'}$ with $\xi_{\mathbf{k'}} = -\xi_{\mathbf{k}}$, so that
both states have the same energy, and their weighting factors in
Eq.~\ref{eq:dos0} add up to unity. Thus, Eq.~(\ref{eq:dos0}) can be
simplified to
\begin{eqnarray*}
\rho(E) &=& \frac{\rho_n(0)}{4\pi} \int_0^{2\pi} d\theta \,
   \int_0^{\hbar \omega_c} d\xi\,\big(\delta(E - |E_{\mathbf{k}+}|) \nonumber\\
&& {} + \delta(E - |E_{\mathbf{k} -}|) \big),
\end{eqnarray*}
where $\rho_n(0)$ is the DOS of the normal state at the Fermi level.
Finally, we obtain
\begin{equation} \label{eq:dos1}
\frac{\rho_e(E)}{\rho_n(0)} = \frac{1}{2\pi}
   \int_0^{2\pi} d\theta \, \tilde{\rho}_e(E,\theta),
\end{equation}
where
\begin{displaymath}
\tilde{\rho}_e(E,\theta) = \frac{1}{2} \Big(
   \frac{|\epsilon_{+}|}{\sqrt{\epsilon_{+}^2 - |\Delta(\mathbf{k}_F)|^2}}
   +\frac{|\epsilon_{-}|}{\sqrt{\epsilon_{-}^2 - |\Delta(\mathbf{k}_F)|^2}}\Big)
   \,,
\end{displaymath}
with $\epsilon_{\sigma} = E - \sigma (\mu_0 B + q v_F
\cos(\theta_{\mathbf{k}} - \theta_{\mathbf{q}}))$. The result is
presented in Fig.~\ref{fig:dos1}.

\section{The conductance characteristics}

\begin{figure}
\includegraphics[width = 0.35\textwidth]{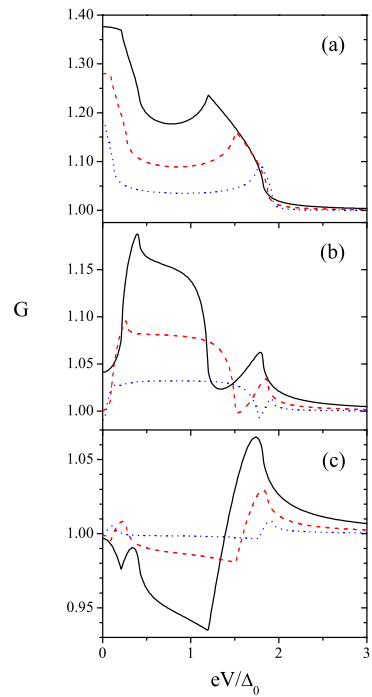}
\caption{\label{fig:btk-s} The normalized conductance vs voltage for
normal-metal/$s$-wave FF superconductor junction: (a) $z=0$, (b) $z=0.5$,
(c) $z=5.0$. Solid, $H=0.704$, $Q = 0.804$; dash, $H = 0.8$, $Q = 0.892$; dot,
$H = 0.9$, $Q = 0.952$.}
\end{figure}

\begin{figure}
\includegraphics[width = 0.35\textwidth]{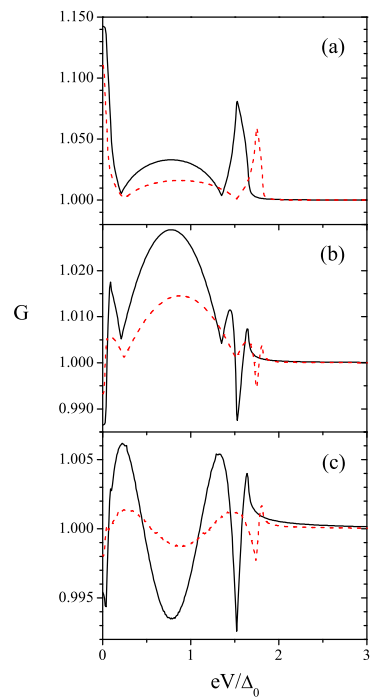}
\caption{\label{fig:btk-d1} Same as Fig.~\ref{fig:btk-s} except it is now for
normal-metal/$d$-wave FF superconductor junction with (100) contact. Solid,
$H = 0.78$, $Q = 0.808$; dash, $H = 0.88$, $Q = 0.9$.}
\end{figure}

\begin{figure}
\includegraphics[width = 0.35\textwidth]{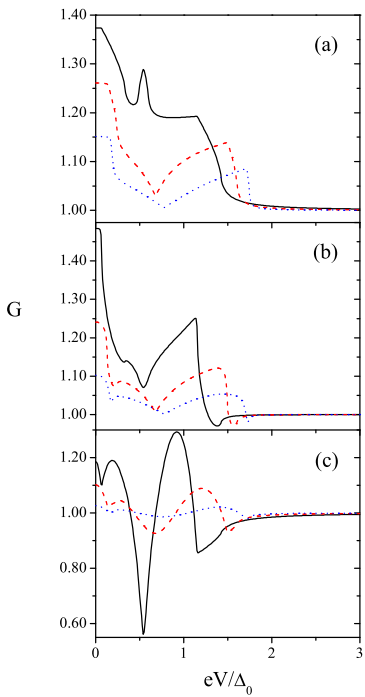}
\caption{\label{fig:btk-d2} Same as Fig.~\ref{fig:btk-s} except it is now for
normal-metal/$d$-wave FF superconductor junction with (110) contact. Solid,
$H = 0.544$, $Q = 0.608$; dash, $H = 0.68$, $Q = 0.816$; dot, $H = 0.776$,
$Q = 0.94$.}
\end{figure}

\begin{figure}
\includegraphics[width = 0.35\textwidth]{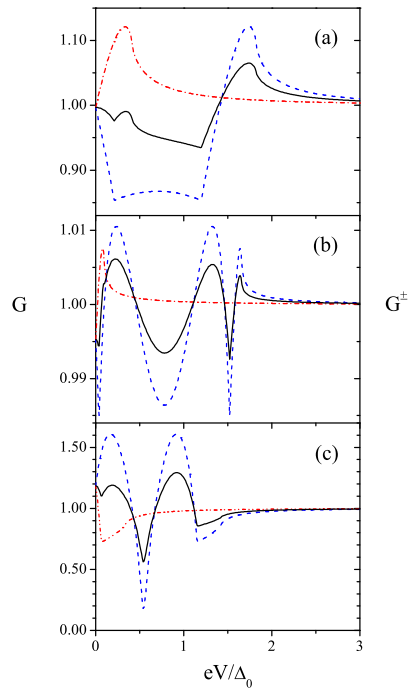}
\caption{\label{fig:spin} The normalized conductance $G$ (solid line) and its two
spin components $G^{\pm}$ (dashed line for spin-up, dash-dotted line for spin-down)
at a large $z = 5.0$. (a) $s$-wave with $H = 0.704$; (b) $d$-wave with
{\bf q} along antinodal direction and $H = 0.78$; (c) $d$-wave with {\bf q}
along nodal direction and $H = 0.544$.}
\end{figure}

\begin{figure}
\includegraphics[width = 0.5\textwidth]{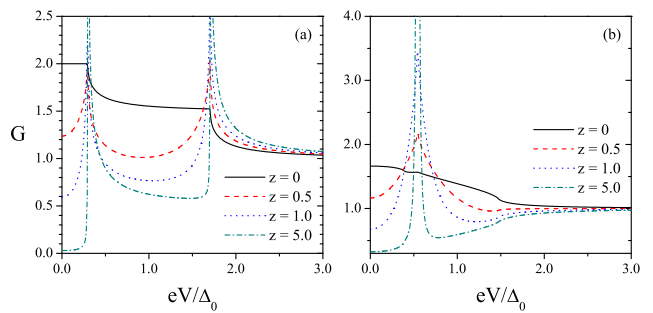}
\caption{\label{fig:btk-non} The normalized conductance vs voltage for
the competing uniform BCS states at the critical fields with different barrier
intensities $z$. (a) $s$-wave superconductor with $H = 0.704$; (b)
$d$-wave superconductor with $H = 0.544$ and (110) contact.}
\end{figure}

In the presence of a normal metal/superconductor interface, to study the QPs,
one needs to solve the Bogoliubov-de Gennes (BdG) equations.~\cite{deGennes}
From the mean field Hamiltonian (\ref{eq:hmltmf}), we obtain
\begin{subequations} \label{eq:BdG}
\begin{eqnarray}
E u_{\sigma}(\mathbf{x}) &=& (\hat{h}_0 + \sigma \mu_0 B) u_{\sigma}(\mathbf{x})
   \nonumber \\
&& + \int d\mathbf{x'} \, \Delta(\mathbf{s, r}) v_{\sigma}(\mathbf{x'}),\\
E v_{\sigma}(\mathbf{x}) &=& -(\hat{h}_0 + \sigma \mu_0 B) v_{\sigma}(\mathbf{x})
   \nonumber \\
&& + \int d\mathbf{x'} \, \Delta^{\ast}(\mathbf{s, r}) u_{\sigma}(\mathbf{x'}),
\end{eqnarray}
\end{subequations}
where $\mathbf{s=x-x'}$, $\mathbf{r=(x+x')}/2$, $\hat{h}_0 =
-\nabla^2/2m + U \delta(x) - \mu$ and $\Delta(\mathbf{s,r}) = \int
d\mathbf{k} \, e^{i\mathbf{k \cdot s}} \tilde{\Delta}(\mathbf{k, r})
e^{i2\mathbf{q \cdot r}}$. Here, {\bf q} is determined by minimizing
the ground-state energy [Eq.~(\ref{eq:FFenergy})]. To avoid
accommodating super/normal current conversion at the N/S interface, which may require a nontrivial modification of the order parameter structure near the interface,
we assume that {\bf q} is parallel to the N/S interface at $x=0$; this choice makes
theoretical analysis simpler and comparison to experiments easier.
Neglecting the proximity effect at the N/S interface, we have
$\tilde{\Delta}(\mathbf{k,r}) = \Delta(\mathbf{k}) \Theta(x)$, where
$\Theta(x)$ is the step function, and $\Delta(\mathbf{k})$ is the
pairing potential in the bulk superconductor. In the WKB
approximation, the BdG equations have the special solutions of the
form,
\begin{equation}
\Big( \begin{array}{c} u_{\sigma} \\ v_{\sigma} \end{array} \Big) =
   e^{i\mathbf{k}_F \cdot \mathbf{x}} \Big( \begin{array}{c}
   e^{i\mathbf{q \cdot x}} \tilde{u}_{\sigma} \\ e^{-i\mathbf{q \cdot x}}
   \tilde{u}_{\sigma}
   \end{array} \Big) \,,
\end{equation}
where $\tilde{u}_{\sigma}$ and $\tilde{v}_{\sigma}$ obey the Andreev
equations
\begin{subequations} \label{eq:Andreev}
\begin{eqnarray}
\epsilon \tilde{u}_{\sigma} &=& -\frac{i(\mathbf{k}_F + \mathbf{q})}{m} \cdot
   \nabla \tilde{u}_{\sigma} + \Delta(\mathbf{k}_F) \Theta(x) \tilde{v}_{\sigma},
   \\
\epsilon \tilde{v}_{\sigma} &=& \frac{i(\mathbf{k}_F - \mathbf{q})}{m} \cdot
   \nabla \tilde{v}_{\sigma} + \Delta^{\ast}(\mathbf{k}_F) \Theta(x)
   \tilde{u}_{\sigma},
\end{eqnarray}
\end{subequations}
where $\epsilon = E -\sigma \mu_0 B - \mathbf{q} \cdot \mathbf{k}_F
/ m$. The eigenenergy $E$ is symmetric about $E = \sigma \mu_0 B +
\mathbf{q} \cdot \mathbf{k}_F / m$ instead of zero as in
Ref.~\onlinecite{BTK}. These equations are similar to those of Ref.
\onlinecite{ZTH2004}, where the authors studied conductance
characteristics in the presence of a supercurrent along the junction
(note that the $q/\Delta^0$ in Ref.~\onlinecite{ZTH2004} is equal to
$Q/2$ of this work). Following the method of
Blonder-Tinkham-Klapwijk~\cite{BTK, tanaka, ZTH2004}, the
normalized tunneling conductance at zero temperature is given by $G
= G_{ns}/G_{nn}$, where $G_{ns}$ and $G_{nn}$ are the tunneling
conductances in the superconducting and normal states, respectively.
Since the QPs of the two spin species are uncoupled, the conductance
will simply be the average over the two spin components:
\begin{displaymath}
G_{ns} = \frac{1}{2}(G_{ns}^{+} + G_{ns}^{-}),
\end{displaymath}
and the same for $G_{nn}$. For simplicity, we will drop the spin
index when there is no ambiguity. The conductances of each spin
species are given by
\begin{equation} \label{eq:g1}
G_{ns} = -\frac{e^2}{\pi} \int_{-\pi/2}^{\pi/2} d\theta \,
   (1+ |a(-E, \theta)|^2 - |b(E, \theta)|^2),
\end{equation}
\vspace{-3ex}
\begin{equation} \label{eq:g2}
G_{nn} = -\frac{e^2}{\pi} \int_{-\pi/2}^{\pi/2} d\theta \,
   (1 - |b(+\infty, \theta)|^2),
\end{equation}
where
\begin{subequations} \label{eq:g3}
\begin{eqnarray}
&& a(E, \theta) = \frac{\cos^2\theta}{\eta_+ (\cos^2\theta + z^2) - \eta_- z^2},
\\
&& b(E, \theta) = -\frac{z(z + i\cos\theta)(\eta_+ - \eta_-)}{\eta_+
   (\cos^2\theta + z^2) - \eta_- z^2}, \\
&& \eta_{\pm} = \frac{\epsilon \pm \sqrt{\epsilon^2 -
   |\Delta(\theta_{\pm})|^2}}{\Delta^{\ast}(\theta_{\pm})}, \\
&& \epsilon = E - \sigma \mu_0 B - \frac{qk_{Fy}}{m}.
\end{eqnarray}
\end{subequations}
Here, $a(E, \theta)$ and $b(E, \theta)$ are the Andreev and normal
reflection coefficients, respectively; $\theta_{\pm} = \theta \pm \alpha$,
$\theta$ is the angle between $\mathbf{k}_F$ and the $+x$ axis,
$\alpha$ is the angle between the antinodal direction and the $+x$
axis for $d$-wave, and zero for $s$-wave, $z=mU/k_F$ is a
dimensionless barrier-strength parameter. We give the normalized
tunneling conductance of the FF states in both $s$- and $d$-wave superconductors
with the arrangement of {\bf q} parallel to the N/S interface in
Figs.~\ref{fig:btk-s}--\ref{fig:btk-d2} , and also show the spin
splitting effect of the tunneling conductance at a large $z = 5.0$
in Fig.~\ref{fig:spin}. In Fig.~\ref{fig:btk-non}, we give the
conductance of the competing uniform BCS states at the critical
fields for both pair types of superconductors. Note that the Zeeman
splitting for the $s$-wave case shown in Fig.~\ref{fig:btk-non}(a)
reproduces the well-known results reviewed in Ref.~\onlinecite{meservey}
and the $z=0$ results given
in Ref.~\onlinecite{melin}.

We start our discussion with an $s$-wave superconductor or a $d$-wave
superconductor with
(100) contact, where there are no ABSs; thus the conductances are
determined by the bulk quasiparticles. We note that, in the
arrangement considered here, due to the effect of pair momentum,
the conductance curves at large $z$ seem to no longer coincide with
the corresponding electron DOS for either case. [Compare
Figs~\ref{fig:btk-s}(c), \ref{fig:btk-d1}(c), and
\ref{fig:btk-d2}(c) with Figs.~\ref{fig:dos1}~(a), (b), and (c).] At
first sight, this appears to be against our intuitive understanding
of what the NSJ conductance at
high-barrier limit (tunneling limit) is supposed to be measuring; we
now resolve this issue below.

\begin{figure}
\includegraphics[width = 0.35\textwidth]{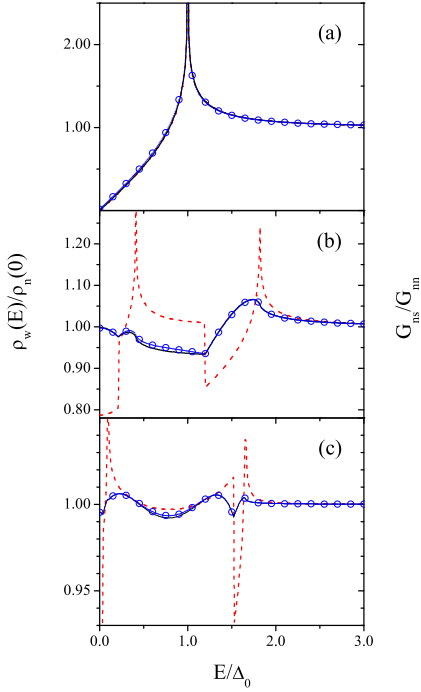}
\caption{\label{fig:dos2} Comparison of the high-$z$(=5.0) junction conductance $G_{ns}$
(circles with a thin dotted line through them) with the un-weighted
(dashed line) and $\cos^2\theta$-weighted (solid line) DOS for (a) an N/($d$-wave
BCS superconductor) junction with (100) contact; (b) an N/($s$-wave FF superconductor) junction,
at $H = 704$; and (c) an N/($d$-wave FF superconductor) junction with (100)
contact and $\bf q$ along (010), at $H = 0.78$.}
\end{figure}

In the arrangement of NSJs considered here, the applied voltage and
the measured current are both along a fixed direction in the
conducting plane that is normal to the N/S interface. Thus we
realize that the junction conductance in the high-barrier-strength
limit is actually measuring a $\cos^2\theta$-weighted DOS. That is,
the QPs of various momenta $\bf k$ on the 2D Fermi surface (i.e.,
circle) do not all make equal contributions to the junction
conductance, but should be weighted by $\cos^2\theta$ where $\theta$
is the angle between $\bf k$ and the current direction $x$. The
weighted DOS [$\rho_w(E)$] measured in the high-$z$-limit junction conductance is
therefore:
\begin{equation} \label{eq:dos2}
\frac{\rho_w(E)}{\rho_n(0)} = \frac{1}{\pi} \int_0^{2\pi}
   d\theta \, \tilde{\rho}_e(E, \theta) |\cos \theta|^2.
\end{equation}
In the uniform BCS state (without involving ABSs), such weighted average simply returns to
the original DOS, i.e., Eq.~(\ref{eq:dos2}) is the same as
Eq.~(\ref{eq:dos1}) in the cases studied here. This is because the
order parameter is isotropic over the momentum space for an $s$-wave
superconductor, while for an N/($d$-wave superconductor) junction with (100) contact, the
order parameter is symmetric about $\theta=\pm\pi/4$ so that the
partial DOS of a fixed direction is also symmetric about the same
line. Because the $\cos^2\theta$ weighting factor adds up to be 1
for two angles that are symmetric about this line, we thus have the
weighted DOS by high-$z$ junction conductance to be again the same
as the un-weighted DOS.~\cite{tanaka} However, the situation is very
different in the FF state, because this kind of symmetry is
broken by the pair momentum which causes a $\theta$-dependent
energy shift, thus the two kinds of DOS are no longer the same. This
is illustrated in Fig.~\ref{fig:dos2}. It is clear from this figure
that the high-$z$ junction conductance measures the
$\cos^2\theta$-weighted DOS, and not the un-weighted DOS in general.
(The slight discrepancy between the tunneling conductance and the weighted DOS is because $z = 5.0$
is still not high enough.)

The most prominent features of the high-$z$ junction conductance of
an N/($d$-wave FF superconductor) with (110) contact are due to the ABSs, which
are the main focus of the present work. To interpret these
features we need to understand how the pair momentum
affects the spectra of the ABSs, which is the subject of the next
section.


\section{Andreev bound states in the d-wave Fulde-Ferrell superconductor with (110) junction}

As has been reviewed in the introduction, for an N/($d$-wave superconductor)
junction with (110) contact, a ZBCP is expected due to the formation
of the ``zero energy Andreev bound states'' (or ``midgap states'')
at the junction interface. Thus for an N/($d$-wave superconductor in an FF
state) junction with (110) contact and {\bf q} along ($1\bar 10$),
we need to understand the effects of $\mu_0 B$ and {\bf q} on the
spectra of ABSs before we can understand the conductance
characteristics for this junction.

In the limit of $z\to\infty$ we can consider a simple model with an
FF superconductor in the region $x \ge 0$, and vacuum or an insulator in the region
$x < 0$, with {\bf q} parallel to the N/S interface. In the superconductor region, from
Eq.~(\ref{eq:Andreev}), we can find the solutions of bound states (which
decays to zero as $x\rightarrow +\infty$) to be of the form,
\begin{displaymath}
\Big( \begin{array}{c} \tilde{u}_{\sigma \pm} \\ \tilde{v}_{\sigma \pm}
\end{array} \Big) = e^{-\gamma x}
\Big( \begin{array}{c} \hat{u}_{\sigma \pm} \\ \hat{v}_{\sigma \pm}
\end{array} \Big)\,,
\end{displaymath}
where
\begin{eqnarray*}
&& \gamma = m \sqrt{|\Delta|^2 - \epsilon^2}/|k_{Fx}|, \\
&& \hat{u}_{\sigma \pm} / \hat{v}_{\sigma \pm} =
   \Delta / (\epsilon \mp i \sqrt{|\Delta|^2 - \epsilon^2})\,,
\end{eqnarray*}
and $\epsilon = E - \sigma \mu_0 B - qk_{Fy}/m$ ($k_{Fx}=k_F \cos\theta$,
$k_{Fy}=k_F \sin\theta$). In order to satisfy
the vanishing wave-function boundary condition at $x=0$, we need to
first superpose the above solutions of opposite $k_x$:
\begin{eqnarray*}
\lefteqn{\psi_{\sigma}(\mathbf{x}) = A e^{i |k_x| x - \gamma x} \left(
   \begin{array}{c} e^{i q y} \hat{u}_{\sigma+} \\ e^{- i q y} \hat{v}_{\sigma+}
   \end{array} \right) {} } \nonumber \\
&& {} + B e^{- i |k_x| x - \gamma x} \left( \begin{array}{c}
   e^{i q y} \hat{u}_{\sigma-} \\ e^{- i q y} \hat{v}_{\sigma-}
   \end{array} \right);
\end{eqnarray*}
imposing the vanishing boundary condition at $x=0$ then yields
\begin{equation} \label{eq:boundbc}
\frac{\hat{u}_{\sigma+}}{\hat{v}_{\sigma+}} =
   \frac{\hat{u}_{\sigma-}}{\hat{v}_{\sigma-}}.
\end{equation}

\begin{figure}
\includegraphics[width = 0.4\textwidth]{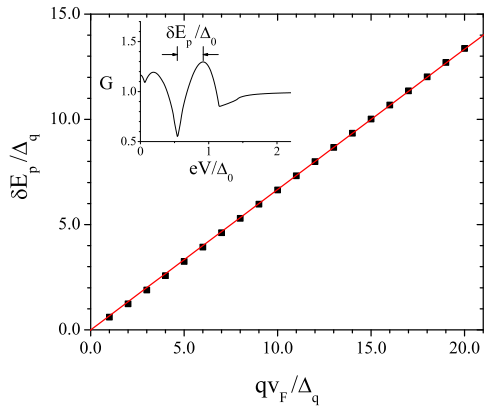}
\caption{\label{fig:peak} The bias voltage difference of the tunneling conductance peak
($E = E_{p+}$) and the dip ($E = \mu_0 B$)
at large $z$(=20), $\delta E_{p} = |E_{p+}
- \mu_0 B|$, as a function of $qv_F$ (both in unit of $\Delta_q$) for
$d$-wave superconductor with (110) contact. The value range of $qv_F/\Delta_q$ is
extended to 0--20, while the physical range is about 1.5--7.7. Solid line
is the linear fit of the numerical data, and its slope is $0.667\pm0.001$. The inset is an
illustration of the voltage difference measured in the experiment.}
\end{figure}

For an $s$-wave superconductor, and for a $d$-wave superconductor with (100) contact,
Eq.~(\ref{eq:boundbc}) yields no ABSs solutions. For a $d$-wave superconductor
with (110) contact, we obtain one solution for each $-k_F < k_{Fy} <
k_F$ at $\epsilon = 0$ or
\begin{equation}
E = \sigma\mu_0 B + qk_{Fy}/m.
\end{equation}
We see that the energies of the ABSs are first split by the Zeeman
energy to $\sigma \mu_0 B$ and then, shifted by an amount proportional to both
the pair momentum {\bf q} and the sine
of the incident angle $\theta$ (which ranges between $-90^{\circ}$
and $+90^{\circ}$); the combined
results of them will lead to a splitting, shifting, and broadening of the
ZBCP, which we now analyze. With (110) contact, the order parameter
is proportional to $\sin 2\theta$, which vanishes at $\theta = 0$,
and $\pm 90^\circ$, implying that near these special values there
are either no ABSs, or their contributions to conductance will be
very weak because these are very loosely bound states; the dominant
contributions come from ABSs with $\theta$ around $\pm \pi/4$ (or
around gap maxima); the energy shift of these states due to pair
momenta are in {\em opposite} direction when the sign of $\theta$ are different.
Thus for sufficiently large $q$, which is the case here, the
junction conductance at positive bias should exhibit two peaks, one
on each side of $\mu_0 B$, and a dip at the Zeeman field energy $E = \mu_0 B$.
(The conductance has a symmetry about zero bias in the approximation
adopted here, so we do not need to consider negative bias.)
For the two peaks, the one on the right side ($E_{p+}$) arises from $0 < \theta
< \pi/2$, whereas the one on the left side ($E_{p-}$) arises from $-\pi/2 < \theta < 0$.
If only spin-up QPs are considered, the two peaks are of equal strength and
equal distance from the dip
as illustrated in the dashed line of Fig.~\ref{fig:spin}(c). When the contributions
from the QPs of both spin species are summed up, the two peaks will not be symmetric
about the dip ($E_{p-}$ will be shifted slightly to the right, but the influence
on $E_{p+}$ and the dip is negligible), and a weak peak at zero bias emerges
(see Fig.~\ref{fig:spin}(c)).
To locate the position of the peak $E_{p+}$, we first notice from Eqs.~\ref{eq:g1}--\ref{eq:g3}
that the bias voltage difference between the peak and the dip,
$\delta E_{p} = |E_{p+} - \mu_0 B|$, should be a function
of $qv_F$ in the high-$z$ limit and vanishes when the pair momentum is zero.
Numerically, therefore we can consider a simplified situation where there exists the
pair momentum without Zeeman field (such in Ref.~\onlinecite{ZTH2004}) and calculate
this difference as a function of the pair momentum. The result is shown in
Fig.~\ref{fig:peak}. By fitting the data to
a straight line through the origin, we obtain
\begin{equation} \label{eq:bspeak-d}
E_{p+} \approx \mu_0 B + \frac{2}{3} q v_F \,.
\end{equation}
Thus, by measuring the bias voltages of the peak and the dip in the
high-$z$ junction conductance with (110) contact, and in particular
the difference between them, we can obtain a good estimate of
{\bf q}. We note that without this pair
momentum, we would have all ABSs at energies $E = \sigma\mu_0 B$,
which would have given rise to one sharp peak only at $eV = \mu_0 B$
in the same conductance plot, as shown in Fig.~\ref{fig:btk-non}(b).
Therefore we conclude that the signature of the FF state is clearly
revealed in the junction conductance characteristics, especially at
high $z$, but the conductance behaviors at low $z$ for all three
cases studied here are also quite novel, since they are quite
different from the corresponding results for uniform $s$- or
$d$-wave superconductors.

\section{Summary and Discussion}

In this paper we have studied the conductance characteristics of a
junction between a normal metal and a superconductor in the Fulde-Ferrell (FF)
state, using the Blonder-Tinkham-Klapwijk formalism. We have studied
both $s$- and $d$-wave cases, and for the latter case, we considered
junctions along both the nodal [(110)] and antinodal [(100)] directions.

The conductance characteristics of a micro-constriction is
presumably easiest to understand in the high tunneling-barrier
limit, when the conductance should give information about DOS of
the superconductor.
In the FF phase, the Zeeman field should split the contributions to
the conductance by the spin-up and -down QPs, i.e., shifting their
contributions by the Zeeman energy in opposite directions. In
addition, for QPs of either spin species, their contributions should
be shifted by an amount proportional to the pair momentum, with a
proportionality constant depending on the cosine of the angle
between its kinetic momentum and the pair momentum. This
proportionality constant can range from a negative maximum to a
positive maximum. Thus we found the effect of {\bf q} to be a
broadening rather than a shift. One would expect similar effects to
occur on the contributions by the ABSs, resulting in the shift and
broadening of the ZBCP. Thus, in the high-barrier-limit, one might
(naively) expect the tunneling spectrum of $d$-wave superconductor
with (110) contact to be composed of broadened ZBCP's centered
around $\pm \mu_0 B$.
In our numerical results, we find instead that the
high-barrier-limit tunneling conductance of $d$-wave superconductors
with (110) contact has a {\em dip} at the Zeeman energy, with one
round peak on each side of it, and also another weak peak at exactly zero
energy. This is quite different from the situation of the BCS state
in the presence of a Zeeman field, in which case one expects the
{\em sharp} ZBCP to be shifted to $\pm \mu_0 B$, as well as the
naive expectation above. The dip and the two round peaks can be
understood as due to the fact that the {\bf q} values appearing here
are so large that they are already beyond the critical value
obtained in Ref.~\onlinecite{ZTH2004}, which studied directly the
current effect on the conductance characteristics in the absence of
a Zeeman field. For such large {\bf q} values, their effect on the
ZBCP in the high-barrier-limit tunneling conductance is to split the
ZBCP into two round peaks with a center dip. It is worth noting that
such high values of {\bf q} are not accessible through direct
application of a supercurrent. The weak peak at zero energy turns
out to be the result of summing up spin-up and -down contributions.
Furthermore, numerical analysis shows that the bias voltage
difference between the dip and the round peak on its right side is
proportional to the pair momentum and thus gives us a simple way to
estimate the pair momentum for $d$-wave superconductor with (110)
contact.

For $s$-wave superconductor and $d$-wave superconductor with (100)
contact, there is no ABS and the conductance is due to contributions
from bulk quasiparticles exclusively. In these cases we have also
found conductance features in the FF superconductors that are very
different from the BCS superconductors.
We found that because of the energy shift due to the pair momentum
(which breaks the spatial symmetries in the original system), the
conductance in the high barrier limit is no longer the same as the
electron DOS; instead, it reflects a {\em directionally-weighted}
DOS. In principle, by comparing the conductance and the bulk
DOS~\cite{maki2004a, vorontsov,mizushima,wang} that are measured by
other means (such as tunneling along the $c$-direction instead of an
in-plane direction discussed here), one can also distinguish between
BCS and FF states.

The FF state studied here is the simplest version of the general
Fulde-Ferrell-Larkin-Ovchinnikov (FFLO) superconductors. Its
simplicity allows for a fairly straightforward calculation of the
conductance characteristics through micro-constriction,~\cite{note1} as well as an
understanding of the results. Study of the conductance
characteristics for general FFLO states is currently under way and
will be reported elsewhere.

\acknowledgments QC and KY were supported by NSF grant No. DMR-0225698. JYTW was supported by NSERC, CFI/OIT, MMO/OCE and the Canadian Institute for Advanced Research.



\begin{thebibliography}{99}
\bibitem{note1} In general the FF state with a single momentum for the order parameter may not
be the lowest energy mean-field state of the system. However it has
the advantage of being the simplest version of the more general FFLO
states, and in particular the Blonder-Tinkham-Klapwijk equations
for conductance can be solved analytically in closed form. It thus
provides a natural starting point, and we expect the qualitative
features of our results to be robust.
\end{thebibliography}
\end{document}